\documentclass[superscriptaddress,secnumarabic,nobibnotes,aps,prd,showkeys,showpacs,nofootinbib,onecolumn,a4paper]{revtex4}
\usepackage{graphicx}
\usepackage{epsf}
\usepackage{bm}
\usepackage{amsmath}
\usepackage{amsfonts}
\usepackage{amssymb}
\usepackage{epstopdf}
\usepackage{hyperref}
\usepackage{color,xcolor}

\usepackage{soul}
\soulregister{\cite}7
\soulregister{\ref}7
\hypersetup{colorlinks=true,linkcolor=blue,anchorcolor=black,citecolor=blue}

\begin{document}
	\title{ Cosmological analysis of a viable $f(R)$ gravity model }
	\author{Siqi He}
	\affiliation{Department of Physics, Liaoning Normal University, Dalian 116029, China}
	\author{Weiqiang Yang}
	\email{d11102004@163.com}

	\affiliation{Department of Physics, Liaoning Normal University, Dalian 116029, China}
	\author{Hangyi Jin}

	\begin{abstract}
		
	Since viable $f(R)$ gravity models must reconcile early-universe inflation with late-time acceleration, we specifically study the dynamical behavior of such a theory during the matter-dominated to dark-energy-dominated transition epoch. By using $y_{H}(z)$ versus  $z$ and the Hubble parameter, we solved the field equations. After appropriately choosing appropriate parameter  values , we plotted a series of images. We mentioned that their current values are similar to latest observations data  and $\Lambda$CDM-model values. Furthermore,  we plotted the fitting of the distance modulus about this model using SN \uppercase\expandafter{\romannumeral1}a observation data. Therefore we find that the $f(R)$ gravity model is consistent with  the SN \uppercase\expandafter{\romannumeral1}a data, meanwhile, explains the late-stage acceleration of the Universe.  Finally,  we used various diagnostic tools including $\left\lbrace r, s\right\rbrace $, $\left\lbrace r, q\right\rbrace $, $w_{D}-w'_{D}$ plane, growth rate analysis,  statefinder hierarchy and  $Om(z)$-diagnostic to evaluate the observational viability of our model, we perform a systematic comparison with the standard $\Lambda$CDM. We found that evolutionary images can be clearly distinguished this model from the $\Lambda$CDM.
	\end{abstract}

	\maketitle
	\date{\today}
	
\section{Introduction}

	In 1916, Einstein discovered the general theory of relativity, ingeniously introducing a constant term into his field equations and publishing the first cosmological solution, which studied the universe as a whole and laid a theoretical foundation in the development history of cosmology. In 1929, Edwin Hubble made the groundbreaking observation that the universe is expanding-a discovery that fundamentally transformed our understanding of cosmology. The phenomenon of redshift revealed that galaxies were moving \label{away from us, and the greater the distance of a galaxy,  the faster its recession speed. The discovery of Hubble's law completely changed people's understanding of the universe and laid the foundation for} modern cosmology.
	In 1998, Riess A. G., Schmidt B. P., and others, as well as Perlmutter and his team in 1999, both supernova research groups, coincidentally used Type Ia supernovae as "candle light sources" for observation. From the luminosity measurements of SN~Ia\cite{SupernovaCosmologyProject:1998vns,SupernovaSearchTeam:1998fmf}, physicists obtained a wealth of astrophysical information, as well as the cosmic microwave background (CMB) observations \cite{WMAP:2003elm,WMAP:2006bqn}  and analysis of the large-scale structure (LSS)\cite{P.J.E. Peebles:2003,E.J. Copeland:2006,Frieman:2008,M.Li:2011}. This further confirmed the indisputable fact of the accelerating expansion of the universe, where the repulsive force effect, rather than the gravitational effect in universal gravitation, plays a dominant role in the evolution of the universe. Meanwhile,  dark energy as the biggest unsolved mysteries  was proposed by several research groups, which was considered as an energy form with negative pressure. It is the $\Lambda$CDM model that provides a successful fit to the observational data. In theory, cosmologists propose various dark energy models.  The literature shows that a number of models have been put forward. These models regard scalar fields as an alternative way to explain dark energy, and quintessence, k-essence, phantom, tachyon, dilation fields, and holographic dark energy are among them\cite{R.R.Caldwell:2002,A.Bouali:2021,M.Malquarti:2003,Emmanuel N:2020,Copeland:2006wr,T.Chiba:2000,S.Wang:2017,F.Bargach:2021,M.Belkacemi:2020,M.Li:2004}.

	In contrast, in the second approach, rather than introducing something on the right-hand side, the left-hand side of the Einstein equation is modified, which results in a modified gravitational theory.\cite{Caldwell:2009ix,K.Bamba:2012} In the last years, these theories have been utilized as a set of proposals to explain, besides other cosmological and astrophysical phenomena, the observed late-stage acceleration of the Universe. In these situations, dark energy or new forms of matter are not required to be considered for explaining the late - time acceleration. Researchers have widely studied these functions as candidate replacements for the theory of General Relativity(GR)\cite{Buchdahl:1970ldb,Starobinsky:1980te,Jamil:2011ptc,Alvarenga:2012bt,Hu:2007nk,Lovelock:1971yv,Nojiri:2005jg,Chiba:2003ir}.
	
	In this paper, we propose a novel $f(R)$ gravity framework capable of simultaneously describing both the primordial inflationary phase and the current dark energy-dominated epoch within a unified theoretical structure.\cite{Shin'ichi Nojiri:2003,Santos:2007bs,Basilakos:2013nfa,Salvatore Capozziello:2019,Hwang:2001pu,Guido Cognola:2005,Bamba:2012qi,Shin'ichi Nojiri:2006,S. Capozziello:2008,DeFelice:2010aj,Ignacy Sawicki:2007,Shin'ichi Nojiri:2007,Song:2007,E. Elizalde:2012,Koivisto:2007sq,Cognola:2008,Elizalde:2010ts,Nojiri:2007cq,Nojiri:2006ri,Nojiri:2010wj,Nojiri:2017ncd,Nojiri:2019fft,Odintsov:2015gba,Odintsov:2017tbc,Odintsov:2018uaw,Odintsov:2019evb,Odintsov:2019mlf,Odintsov:2020nwm,Oikonomou:2020qah,Odintsov:2026fyc,Odintsov:2025jfq,Odintsov:2024woi,Oikonomou:2020oex,S.D.Odintsov:2020,S.D.Odintsov:2019,Oikonomou:2013rba,Oikonomou:2022wuk,Nojiri:2006gh,Pal:2025zep,Pal:2026hkq,Bezrukov:2007ep} A more general function $f(R)$ ,which is considered in the EH action about R. Whereas  GR exhibits limitations in certain cosmological regimes, an intensive analysis of $f(R)$ gravity demonstrates its enhanced capacity to address these theoretical challenges.\cite{Nojiri:2006gh}.  Provided that specific consistency conditions and multi-scale constraints are satisfied, the exact definition of the function $f(R)$ can be determined, which require the model to simultaneously pass cosmological observations. Two particularly well-studied $f(R)$ models in the literature that satisfy these requirements are the Hu-Sawicki model \cite{Hu:2007nk} and the Starobinsky model \cite{A.A. Starobinsky:2007}. Although initially, these models were marketed as those where the cosmological constant is not incorporated within the $f(R)$ framework, setting them apart from the $\Lambda$CDM version where $f(R)$ , it has been demonstrated that these models can be arbitrarily close to $\Lambda$CDM and behaves extremely well on large scales. Following this footprints of the article, we will introduce an $f(R)$ gravity model characterized by the $R^{2}$ formulation\cite{Bezrukov:2007ep}.
	
	In our paper, we shall consider a polynomial format $f(R)$ model.\cite{Oikonomou:2022wuk}. In Sect.2, we will briefly present the fundamental features of $f(R)$ gravity model encompassing the derivation of field equations via a statefinder function and their subsequent numerical solution. Then we mainly focus in the late-stage cosmological analysis of the model including evolution  of cosmological parameters  in Sect.3. The parameter behave between our model and  $\Lambda$CDM follows a similar trajectory when considering the statefinders. To compare the $f(R)$ gravity with the $\Lambda$CDM, several cosmological diagnostics and observational tests are typically used to assess their viability and agreement with empirical evidence.  We consider five cosmological parameters in the third section. In Sect.4,we give a brief introduction of polynomial format $f(R)$ model.  In Sect. 5,  We consider the SN \uppercase\expandafter{\romannumeral1}a observational dataset and proceed to compare the theoretical outcomes with those of the $\Lambda$CDM model, observing a satisfactory consistency between them and the SN \uppercase\expandafter{\romannumeral1}a data. Then in the most important part, we propose various diagnostic tools to the $f(R)$ gravity model, including the statefinder diagnostic\cite{J. F. Zhang:2014,Alam:2003sc,Zhang:2009qa,Sahni:2002fz,Sahni:2008xx}, growth rate of perturbations, statefinder hierarchy\cite{M. Arabsalmani:2011,Omar Enkhili:2024,D.Mhamdi:2024}, analysis by the $w_{D}-w'_{D}$ plane\cite{Guo:2006pc} and $Om$ diagnostic\cite{Oliveros:2023ewl}. In the process of geometric diagnosis, this model is distinguished from $\Lambda$CDM model and approaches it in the future.
	  Finally, our conclusions are exposed in Sect.7.

	\section {BRIEF REVIEW OF $f(R)$ GRAVITY THEORY }

	\label{section2}
     The fundamental action defining $F(R)$ gravity takes the form\cite{Oikonomou:2022wuk}
	\begin{equation}
		\label{1}
		S=\int d^{4}x\sqrt{-g}(\frac{f(R)}{2\kappa^{2}}+\mathcal{L}_{m}),
	\end{equation}
    here $g$ represents metric tensor $g^{\mu\nu}$. $G$ is Newton’s constant as  well as $M_{pl}$ is the reduced Planck mass. The symbol $\mathcal{L}_{m}$  represents the matter fluid Lagrangian density; whereas $f(R)$ denotes an arbitrary functional form of the Ricci scalar . Varying the action with respect to the metric tensor $g_{\mu\nu}$ gives the field equations
    \begin{equation}
    	\label{2}
    	2f_{R}(R)R_{\mu\nu}-g_{\mu\nu}f(R)+2(g_{\mu\nu}\Box-\nabla_{\mu}\nabla_{\nu}) f_{R}(R)=2\kappa^{2}T_{\mu\nu},
    \end{equation}
    here $f_{R}(R)=\dfrac{df(R)}{dR}$, The covariant derivative, denoted by $\nabla_{\mu}$, is constructed from the Levi-Civita connection of the metric. The d'Alembertian is then given in terms of this derivative as $\Box\equiv\nabla^{\mu}\nabla_{\mu}$. In addition, $T_{\mu\nu}=-\dfrac{2}{\sqrt{-g}}\dfrac{\delta{\mathcal{L}_{m}}}{\delta{g^{\mu\nu}}}$ is an energy-momentum tensor of matter.

     By varying the action (\ref{1}), the field equations take the forms are given by
    \begin{equation}
    	\label{3}
    	3f_{R}H^{2}=\kappa^{2} \rho_{m}+\dfrac{f_{R}R-F}{2}-3H\dot{f_{R}},
    \end{equation}
    \begin{equation}
    	\label{4}
    	-2f_{R}\dot{H}=\kappa^{2}(\rho_{m}+P_{m})+\ddot{F}-H\dot{F}.
    \end{equation}   
    where $H=\dfrac{\dot{a}}{a}$,  ($\dot{}$) represents the rate of change with respect to cosmic time.  meanwhile, $\rho_{m}$ and $P_{m}$ are the energy density and the pressure of perfect fluid matter, respectively. We can obtain from  Eqs. (\ref{4}) and (\ref{5}) as follows
    \begin{equation}
    	\label{5}
    	H^{2}=\frac{\kappa^{2}}{3}(\rho_{m}+\rho_{DE}+\rho_{r}).
    \end{equation}
   
    We additionally posit that the cosmological fluid comprises three distinct components. We can derive from the continuity equation that,
    \begin{equation}
    	\label{6}
    	\dot{\rho_{DE}}+3H(1+w_{DE})\rho_{DE}=0
    \end{equation}
    where $w_{DE}=\frac{P_{DE}}{\rho_{DE}}$ , $p_{DE}$ is pressure for the dark energy.
    
    It follows that the dark energy component is geometric in origin, a property revealed after performing appropriate manipulations on Eqs.(\ref{4}) and (\ref{5}). We find the quantities $\rho_{DE}$ and $P_{DE}$ are defined through the Friedmann and Raychaudhuri equations, reads
     \begin{equation}
    	\label{7}
    	\rho_{DE}=\dfrac{f_{R}-f}{2}-3H\dot{f_{R}}+3H^{2}(1-f_{R}),
    \end{equation}
   
    \begin{equation}
    	\label{8}
    	P_{DE}=\ddot{f}+2\dot{H}(f_{R}-1)-H\dot{f}-\rho_{DE}.
    \end{equation}
    According to Ref.\cite{S.D.Odintsov:2020}, we can get $\rho_{m}=\rho_{m}^{(0)}(a^{-3}+\chi a^{-4})$, where $\rho_{m}^{(0)}$ refers to the contemporary value of the cold dark matter density. $\chi=\dfrac{\rho_{r}^{(0)}}{\rho_{m}^{(0)}}\simeq3.1\times10^{-4}$,  $\rho_{r}^{(0)}$ is current radiation density.
    We choose using the redshift $z$ as a dynamical variable to quantify evolution , which is defined as $1+z=\dfrac{1}{a}$.  Then we shall give a introduction of the statefinder function $y_{H}$
    \begin{equation}
    	\label{9}
    	y_{H}\equiv\dfrac{\rho_{DE}}{\rho_{m}^{(0)}}=\dfrac{H^{2}}{m_{s}^{2}}-a^{-3}-\chi a^{-4},
    \end{equation}  
    where $m_{s}^{2}=\dfrac{\kappa^{2}\rho_{m}^{(0)}}{3}=H_{0}^{2}\Omega_{m}=1.37\times 10^{-67}eV^{2}$, which is the mass scale. By using the Eqs.(\ref{4}) and (\ref{9}), we could get\cite{K.Bamba:2012,S.D.Odintsov:2020,Hu:2007nk}
    \begin{equation}
    	\label{10}
    	\dfrac{d^{2}y_{H}}{dz^{2}}+J_{1}\dfrac{dy_{H}}{dz}+J_{2}y_{H}+J_{3}=0,
    \end{equation}
    where the dimensionless function $J_{1} , J_{2} , J_{3}$ are defined as
    \begin{equation}
    	\label{11}
    	J_{1}=(z+1)^{-1}(-\dfrac{1}{(z+1)^{3}+y_{H}+\chi (z+1)^{4}}\dfrac{1-f_{R}}{6m_{s}^{2}f_{RR}}-3),
    \end{equation}
    \begin{equation}
    	\label{12}
    	J_{2}=(z+1)^{-2}(\dfrac{1}{(z+1)^{3}+\chi (z+1)^{4}+y_{H}}\dfrac{2-f_{R}}{3m_{s}^{2}f_{RR}}),
    \end{equation}
    \begin{equation}
    	\label{13}
    	J_{3}=-3(z+1)-\dfrac{(R-f)/(3m_{s}^{2})+(1-f_{R})[(z+1)^{3}+2\chi(z+1)^{4}]}{(z+1)^{2}((z+1)^{3}+\chi(z+1)^{4}+y_{H})}\dfrac{1}{6m_{s}^{2}f_{RR}}.
    \end{equation}
    where $f_{RR}=\dfrac{\partial ^{2}f}{\partial R^{2}}$. Given that the numerical integration of Eq.(\ref{10}) demands physically realistic boundary values, we establish the initial conditions at the formation redshift  $z_{f}=10$.\cite{K.Bamba:2012,S.D.Odintsov:2020}.  Based on the numerical solution for $y_{H}(z)$, key cosmological quantities can be derived, including the Hubble parameter, Ricci scalar, dimensionless dark energy density parameter ($\Omega_{DE}$), the  equation-of-state of dark energy ($w_{DE}$), the effective equation-of-state ($w_{eff}$), and the deceleration parameter ($q$), all expressed in terms of $y_{H}(z)$.  We will discuss the evolution of these parameters in our $f(R)$ gravity model.
    \begin{equation}
    	\label{14}
    	y_{H}(z_{f})=\dfrac{\Lambda}{3m_{s}^{2}}(\dfrac{1+z_{f}}{1000}+1),
    \end{equation}
    \begin{equation}
    	\label{15}
    	\dfrac{dy_{H}(z)}{dz}\bigg|_{z=z_{f}}=\dfrac{1}{1000}\dfrac{\Lambda}{3m_{s}^{2}},
    \end{equation}
    where $z_{f}=10$ and $\Lambda\simeq 11.895\times10^{-67}eV^{2}$.
    
    \section {COSMOLOGICAL PARAMETERS }
    We consider cosmological parameters that can predict and explain the cosmological evolution. Here, through equations, we can get,
    \begin{equation}
    	\label{16}
    	\Omega_{DE}(z)=\dfrac{y_{H}(z)}{(z+1)^{3}+\chi (z+1)^{4}+y_{H}(z)},
    \end{equation}
    Without considering the radiation, the matter density parameter is given by,
    \begin{equation}
    	\label{17}
    	\Omega_{m}(z)=\dfrac{(1+z)^{3}}{(z+1)^{3}+y_{H}(z)+\chi (z+1)^{4}},
    \end{equation}
      Similarly, $w_{DE}$ and $w(z)$ are described as follows,
     \begin{equation}
     	\label{18}
     	w_{DE}(z)=\dfrac{1}{3}(z+1)\dfrac{1}{y_{H}}\dfrac{dy_{H}(z)}{dz}-1,
     \end{equation}
     
     \begin{equation}
     	\label{19}
     	w(z)=-\frac{2}{3}\dfrac{\dot{H}}{H^{2}}-1=\dfrac{2(z+1)H'(z)}{3H(z)}-1,
     \end{equation}
 the deceleration parameter is
 \begin{equation}
 	\label{20}
 	q(z)=-\dfrac{\dot{H}}{H^{2}}-1=-(z+1)\dfrac{H'(z)}{H(z)}-1,
 \end{equation}
 The notation $'$ indicates differentiation with respect to the redshift $z$.

    \section{LATE-STAGE~$f(R)$~GRAVITY~DYNAMICS}
   		We will proceed to exhibit the numerical findings pertaining to the model introduced in the preceding part. Fundamentally, we compute the current values and illustrate the evolution of selected statefinders at low redshifts. We assume the polynomial type function as follows,\cite{Oikonomou:2022wuk}
    \begin{equation}
    	\label{21}
    	f(R)=R+\dfrac{R^{2}}{M^{2}}-\lambda\frac{c-(\mu^{2}/R)^{\eta}}{d+(\mu^{2}/R)^{\eta}},
    \end{equation}

\begin{figure}[ptb]
	\centering
	\includegraphics[width=0.34\textwidth]{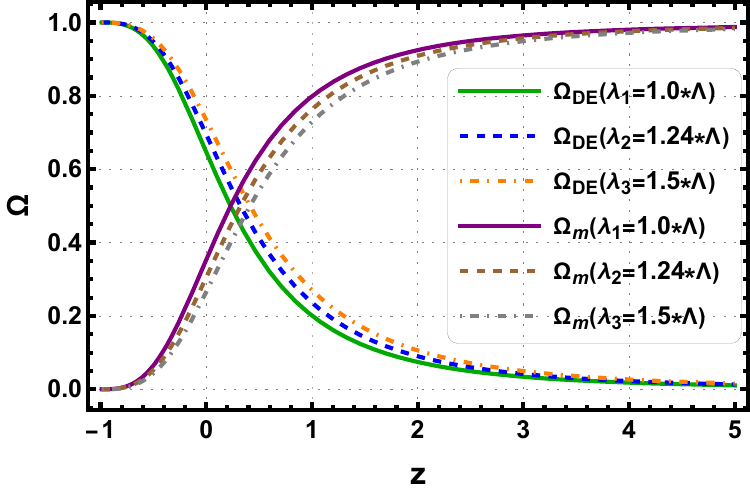}~~~~
	\includegraphics[width=0.34\textwidth]{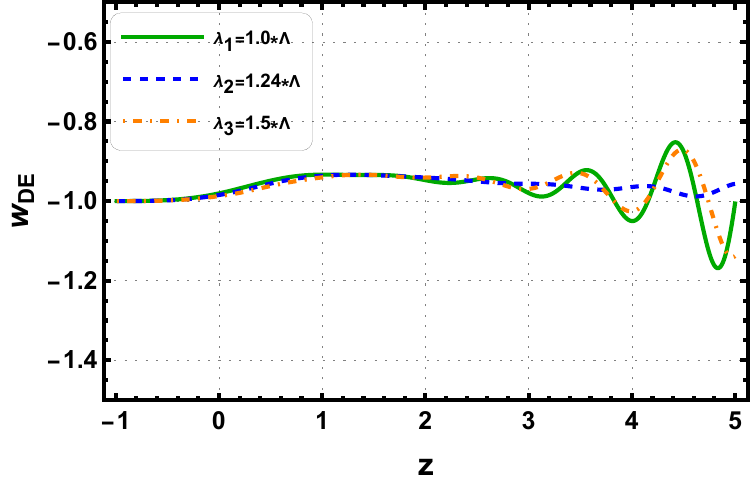}
	\caption{\small{When $\lambda$ takes different values ($\lambda_{1} =1.0\Lambda$, $\lambda_{2} = 1.24\Lambda$, $\lambda_{3} =1.5\Lambda$), $\eta=0.09$ and $\mu^{2}=1.37\times10^{-67}eV^{2}$,  the density parameters $\Omega_{DE}$ and $\Omega_{m}$ (left plot) versus $z$ and $w_{DE}(z)$ (right plot) for the polynomial type  $f(R)$																										 model of Eqs.(\ref{17})-(\ref{18}).}}
	\label{fig:fig1}
\end{figure}

\begin{figure}[ptb]
	\centering
	\includegraphics[width=0.34\textwidth]{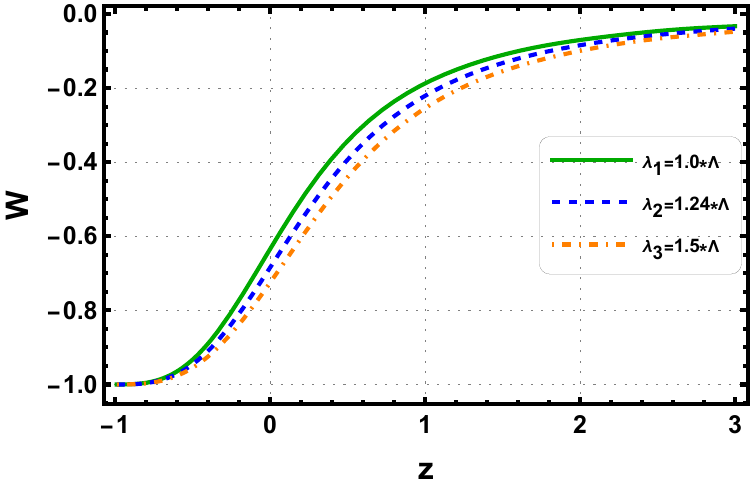}~~~~
	\includegraphics[width=0.35\textwidth]{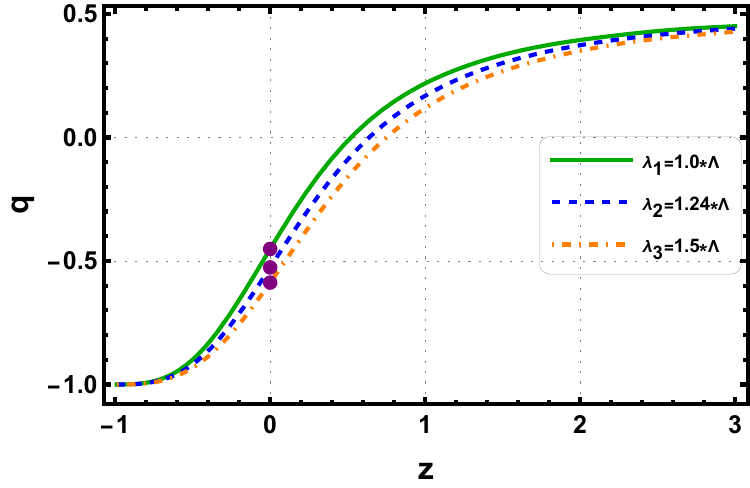}
	\caption{\small{When $\lambda$ takes different values ($\lambda_{1} =1.0\Lambda$, $\lambda_{2} = 1.24\Lambda$, $\lambda_{3} = 1.5\Lambda$), $\eta=0.09$ and $\mu^{2}=1.37\times10^{-67}eV^{2}$, $w(z)$ and $q$are plotted against  $z$ within the polynomial type  $f(R)$																										 model  of Eqs.(\ref{19}) and (\ref{20}).}}
	\label{fig:fig2}
\end{figure}

   In this context, $\lambda$ and $\mu^2$ represent dimensionless parameters (in natural units) with units of $\text{eV}^2$, corresponding to a mass dimension of 2 and $c$, $d$, $\eta$ are dimensionless free parameters\cite{Oikonomou:2022wuk}. We set different value equal to $\lambda_{1} =1.0\Lambda$, $\lambda_{2} = 1.24\Lambda$, $\lambda_{3} = 1.5\Lambda$, $c = 5.5$, $d=3$, $\eta=0.09$ and $\mu =m_{s}$, then we get a viable phenomenology.
    It is worth mentioning that we have appended an $R^{2}/M^{2}$ term to the function in this discussion, where $M=3.04375\times10^{22}eV$.

 In Fig.\ref{fig:fig1}, the density parameter $\Omega_{DE}$, $\Omega_{m}$ and the image of $w_{DE}$ are plotted versus the redshift $z$, which perfectly shows the transformation from the dark matter stage to the dark energy epoch. Without radiation, satisfies $\Omega_{DE}+\Omega_{m}=1$. In the future scenario, they satisfy $w_{DE}=-1$, suggesting that the future fluid behavior aligns precisely with that of the $\Lambda$CDM.  In Fig.\ref{fig:fig2}, for three values , the present value of $w(z)$ fluctuates around $-0.7$, which shows that matter-dominated era comes before and is then replaced by a dark-energy-dominated era. First, it describes how the universe shifts from decelerated expansion to accelerated expansion. Based on the present value $q_{0}<0$ of the deceleration parameter in the second picture, it becomes evident that the present-day universe is in the accelerated expansion era.

   	\begin{figure}[ptb]
   	\centering
   	\includegraphics[width=0.34\textwidth]{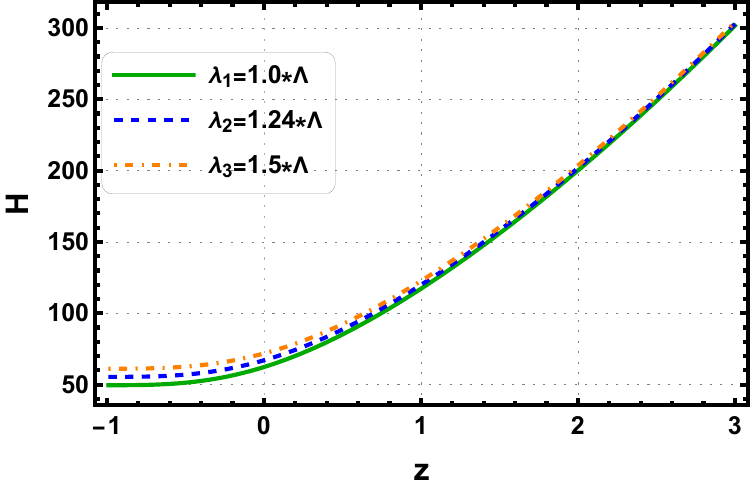}~~~~
   	\includegraphics[width=0.34\textwidth]{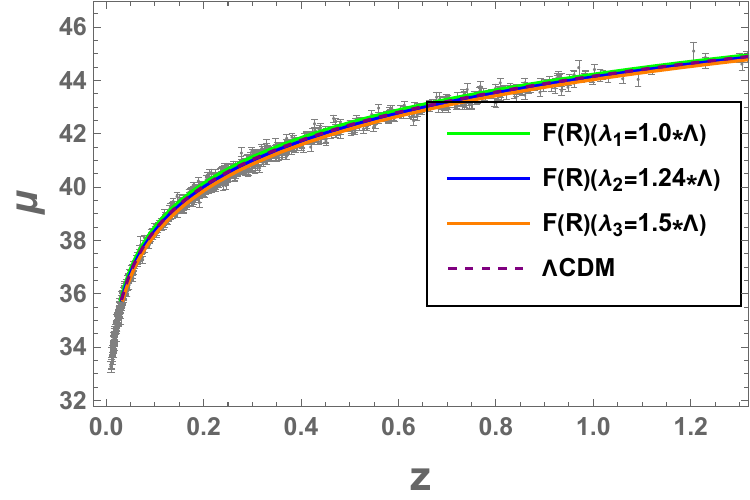}
   	\caption{\small{When $\lambda$ takes different values ($\lambda_{1} =1.0\Lambda$, $\lambda_{2} = 1.24\Lambda$, $\lambda_{3} = 1.5\Lambda$), $\eta=0.09$ and $\mu^{2}=1.37\times10^{-67}eV^{2}$,  $H(z)$ (left plot) and  $\mu$  comparing with $1048$ SN \uppercase\expandafter{\romannumeral1}a observation data points(gray bars) are plotted versus $z$ within for the polynomial type  $f(R)$ model of  Eqs.(\ref{24}) and (\ref{27}).}}
   	\label{fig:fig3}
   	
   \end{figure}

     \section{ SUPERNOVAE OBSERVATIONS WITH $F(R)$ MODEL}

    The luminosity distance $d_{L}$ can be observationally measured by making use of the Supernova type Ia(SN \uppercase\expandafter{\romannumeral1}a), which is a kind of \textquotedblleft standard candle\textquotedblright. In SN \uppercase\expandafter{\romannumeral1}a observations, the luminosity distance $d_{L}$ is employed with the specific aim of connecting the supernova luminosity to the expansion rate of the Universe.  To provide a more impartial and clearly observable perspective on cosmological behavior, we contrast SN \uppercase\expandafter{\romannumeral1}a observation data with theoretical date.
    \begin{equation}
    	\label{24}
    	H(z)=2m_{s}\sqrt{(z+1)^{3}(1+(1+z)\chi)+y_{H}(z)}.
    \end{equation}  

   $H(z)$ has been plotted in the second picture. Using the latest Planck data\cite{Planck:2018vyg}, we can obtain $H_{0}=67.4~ km ~s^{-1}~Mpc^{-1}$, which is roughly consistent with our model. The relationship between $d_{L}$ versus $z$ is able to be described as
    \begin{equation}
    	\label{25}
    	d_{L}=a_{0}(1+z)r.
    \end{equation}	
    The relationship between $d_{L}$ and $H(z)$ is given as follows:
    \begin{equation}
    	\label{26}
    	d_{L}=\dfrac{c}{H_{0}}(1+z)\int_{0}^{z}\dfrac{dz}{E(z)}
    \end{equation}

    The absolute magnitude $M$ of an astronomical object is related to its apparent magnitude $m_{b}$ and luminosity distance $d_{L}$, while the distance modulus $\mu$ is according  to $d_{L}$through the expression: 
    
    \begin{equation}
    	\label{27}
    	\mu =m_{b}-M=5log_{10}\dfrac{d_{L}}{Mpc}+25.
    \end{equation}

    The expansion history $H(z)$ is obtained by solving the background evolution equations numerically. The resulting best-fit curve for $\mu$ as a function of $z$ in the polynomial type  $f(R)$ framework is presented in Fig. \ref{fig:fig3}. Based on the likelihood analysis of the SN \uppercase\expandafter{\romannumeral1}a data points (gray error bars) from\cite{D.M.Scolnic:2018}. As shown in the figure, the theoretical curves for both cases align with the majority of observation points. This close agreement indicates that the $f(R)$ gravity model is compatible with current cosmological observations and provides a consistent description of the universe's accelerated expansion.

    \begin{center}
   	{\small Table.~1~  we list the current values $(z=0)$ of these parameters of  $f(R)$ model ($\lambda_{1} =1.0\Lambda$, $\lambda_{2} = 1.24\Lambda$, $\lambda_{3} = 1.5\Lambda$) and $\Lambda$CDM model  in comparison to the Planck 2018 or SN Ia using Eqs.(\ref{18})-(\ref{20})}
   	\begin{tabular}{|c|c|c|c|c|c|}
   		\hline
   		~~&\multicolumn {3}{|c|}{$F(R)$}   & 
   		\multicolumn {1}{|c|}{$\Lambda$CDM} & 
   		\multicolumn {1}{|c|}{$Planck ~2018 ~or~ SN ~Ia$} \\
   		\hline
   		$Parameters$ &$\lambda_{1}$&$\lambda_{2}$&$\lambda_{3}$&-&-\\
   		\hline
   		$\Omega_{DE}^{(0)}$&0.6467&0.6949&0.7344&0.685&0.6847$\pm$0.0073\\
   		\hline
   		$w_{de}^{(0)}$&-0.981&-0.985&-0.987&-1&$-1.03\pm0.03$ \\
   		\hline
   		$q_{0}$&-0.451&-0.526&-0.587&-0.528&$-0.53_{-0.13}^{+0.17}(SN ~Ia)$ \\
   		\hline
   		$w_{0}$&-0.634&-0.684&-0.725&-0.685&- \\
   		\hline
   			$H_{0}$&62.30&67.02&71.83&67.4&$67.4\pm0.5$\\
   		\hline
   	\end{tabular}
   \end{center}

\section{COSMOLOGICAL GEOMETRICAL DIAGNOSTICS}

    Through the above analysis, we notice that $H(z)$ and  $q$ are not obvious for distinguishing between the $f(R)$ model and $\Lambda$CDM . So in this part, we will introduce four diagnostic tools in order to compare them.

    \subsection{Statefinder analysis}
    
    The statefinder parameters $\left\lbrace r, s \right\rbrace$, which are among the most prominent geometric diagnostics in cosmology, were first proposed by Sahni \textit{et al.} \cite{V.Sahni:2008}, which can be written mathematically as follows:
    \begin{equation}
    	\label{28}
    	r\equiv\dfrac{\dddot{a}}{aH^{3}}
    \end{equation}
    \begin{equation}
    	\label{29}
    	s\equiv\dfrac{r-1}{3q-\dfrac{3}{2}}
    \end{equation}
    here, the overdot signifies differentiation with respect to cosmic time, and the triple derivative $\dddot{a}$ corresponds to the third-order temporal derivative of the cosmological scale factor.
    
    \begin{figure}[ptb]
    	\centering
    	\includegraphics[width=0.34\textwidth]{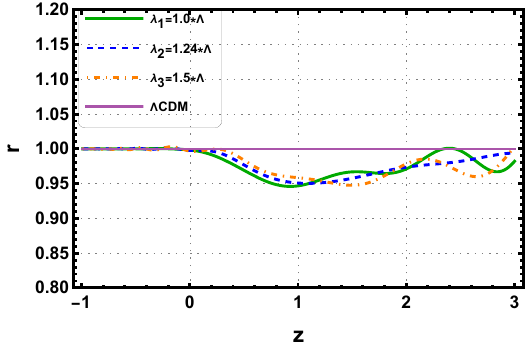}
    	\includegraphics[width=0.34\textwidth]{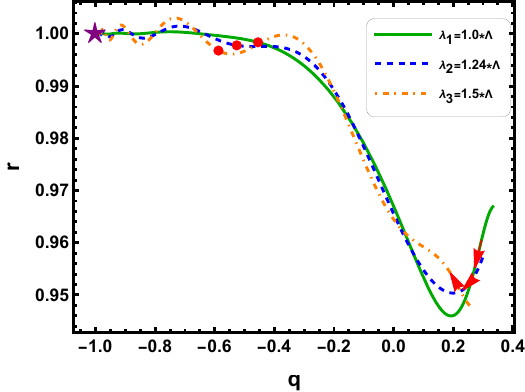}
    	\caption{\small{When $\lambda$ takes different values ($\lambda_{1} =1.0\Lambda$, $\lambda_{2} = 1.24\Lambda$, $\lambda_{3} = 1.5\Lambda$), $\eta=0.09$ and $\mu^{2}=1.37\times10^{-67}eV^{2}$, the evolutions of statfinder parameters for the polynomial type  $f(R)$																										 model.}}
    	\label{fig:fig4}
    \end{figure}	
   
    In Fig.\ref{fig:fig4},  by taking different parameter values $\lambda$, we notice that the trajectories are different, however, for $\lambda_{1} =1.0\Lambda$, $\lambda_{2} = 1.24\Lambda$ and $\lambda_{3} = 1.5\Lambda$, their $\left\lbrace r,s\right\rbrace $ values will eventually approach  $(1,0)$ , which indicates that the trend will approach the evolution of $\Lambda$CDM. We are able to obtain that the $\left\lbrace r,q\right\rbrace $ trajectory plots all end at $r=1,q=-1$ in the right side in the late-stage universe, which can better distinguish them in the low-redshift region. We are able to explicitly observe that $r-s$ plane  and  $r-q$ plane possess a dual - distinguishing ability. Not only can they differentiate the $f(R)$ model from the $\Lambda$CDM , but they can also distinguish between different $\lambda$ within the same model form.

    \subsection{ Statefinder hierarchy and growth rate of perturbations}

    In the first discussion, we we derive the general formulations of the statefinder hierarchy and present their specific expressions in terms of  $H$, $q$  as well as  $z$. The second part describes a brief introduction of the growth rate of perturbations.
   
   \subsubsection{The statefinder hierarchy}
  In order to describe cosmic expansion dynamics, the scale factor $a(t)$ can be represented as a Taylor expansion around the current cosmic time $t_0$. \cite{M. Arabsalmani:2011}:
   \begin{equation}
   	\label{30}
   	\frac{a(t)}{a_{0}}=1+\sum^{\infty}_{n=1}\frac{A_{n}(t_{0})}{n!}((t-t_{0})H_{0})^{n},
   \end{equation}
  The first-order statefinder parameter of hierarchy level $n$ is defined as:
   \begin{equation}
   	\label{31}
   	A_{n}=\frac{a(t)^{(n)}}{a(t)H^{n}},        n\in N.
   \end{equation}

   \begin{figure}[ptb]
   	\centering
   	\includegraphics[width=0.32\textwidth]{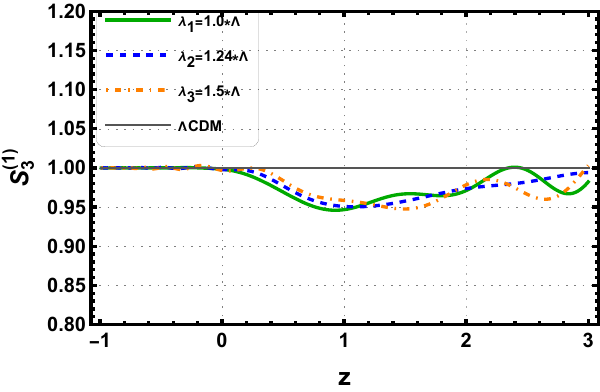}
   	\includegraphics[width=0.32\textwidth]{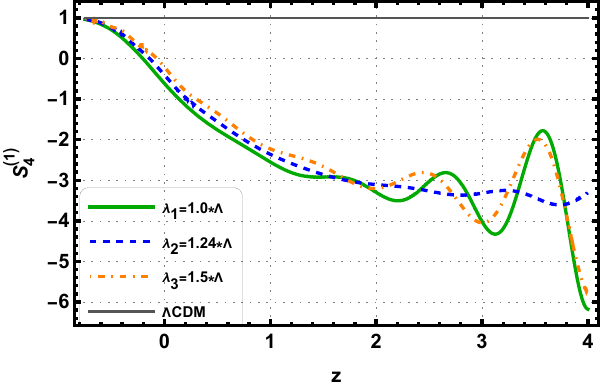}
   
   	\caption{\small{When $\lambda$ takes different values ($\lambda_{1} =1.0\Lambda$, $\lambda_{2} = 1.24\Lambda$, $\lambda_{3} = 1.5\Lambda$), $\eta=0.09$ and $\mu^{2}=1.37\times10^{-67}eV^{2}$,  the evolutions of $S_{3}^{(1)}$ and $S_{4}^{(1)}$ versus  $z$ for the polynomial type  $f(R)$																										 model.}}
   	\label{fig:fig5}
   \end{figure}

 \begin{figure}[ptb]
	\centering
	\includegraphics[width=0.32\textwidth]{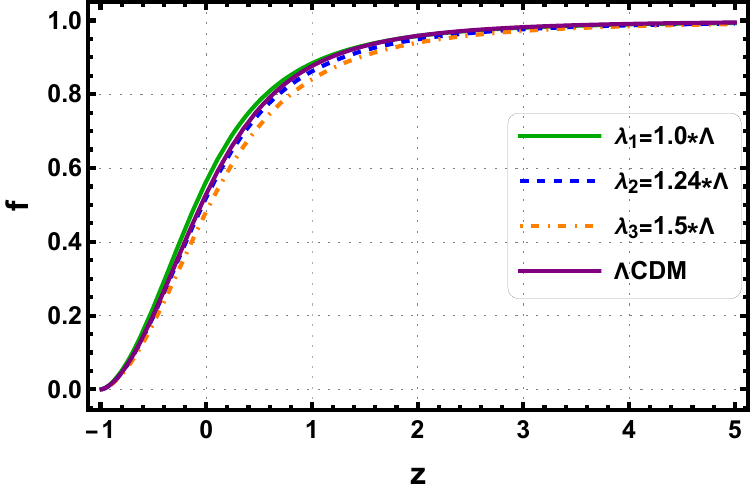}
	\includegraphics[width=0.32\textwidth]{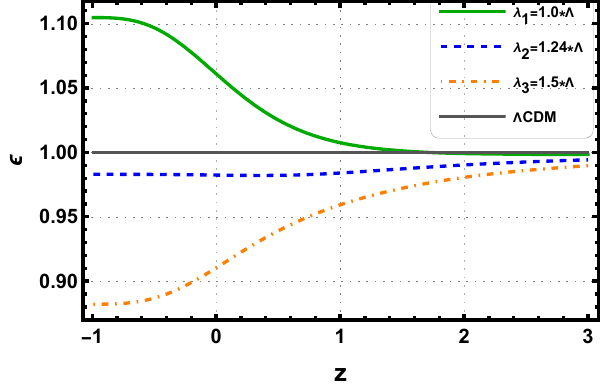}
	\caption{\small{When $\lambda$ takes different values ($\lambda_{1} =1.0\Lambda$, $\lambda_{2} = 1.24\Lambda$, $\lambda_{3} = 1.5\Lambda$), $\eta=0.09$ and $\mu^{2}=1.37\times10^{-67}eV^{2}$,  the evolutions of  growth rate $f$ and  $\epsilon$ versus  $z$ for the polynomial type  $f(R)$																										 model.}}
	\label{fig:fig6}
\end{figure}
   \begin{figure}[ptb]
   	\centering
   	\includegraphics[width=0.34\textwidth]{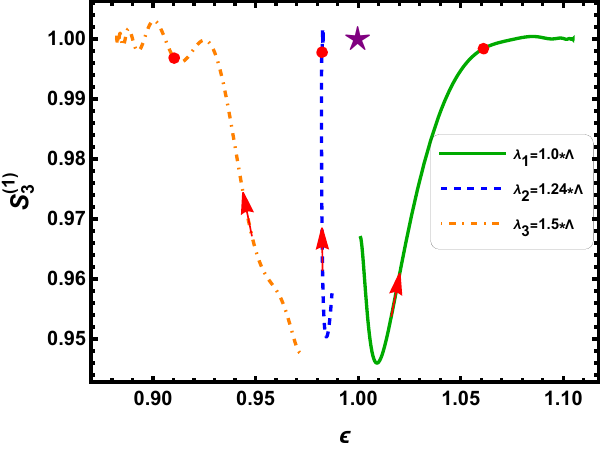}
   	\includegraphics[width=0.33\textwidth]{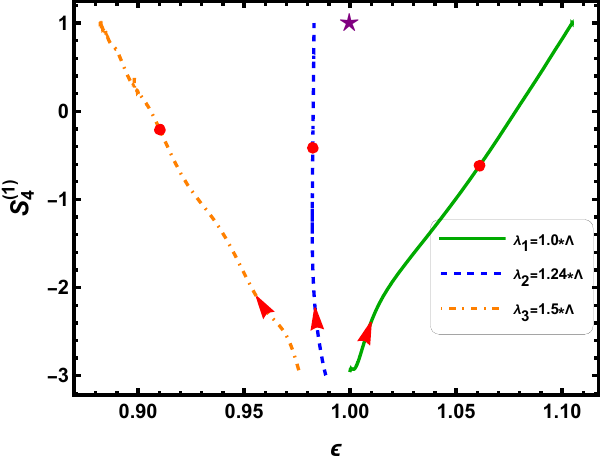}
   	\caption{\small{When $\lambda$ takes different values ($\lambda_{1} =1.0\Lambda$, $\lambda_{2} = 1.24\Lambda$, $\lambda_{3} = 1.5\Lambda$), $\eta=0.09$ and $\mu^{2}=1.37\times10^{-67}eV^{2}$.  The composite null diagnostics $S_{3}^{(1)}$ and $S_{4}^{(1)}$ are plotted for the $f(R)$ model. }}
   	\label{fig:fig7}
   \end{figure}
   In the $\Lambda$CDM framework, the parameter $S_{n}^{(1)}$ takes a value of unity, providing a diagnostic criterion for discriminating between alternative cosmological models. Meanwhile, $n\geqslant3$, which can be written as
   \begin{equation}
   	\label{38}
   	S_{3}^{(1)}=A_{3}
   \end{equation}
   \begin{equation}
   	\label{39}
   	S_{4}^{(1)}=A_{4}+3(1+q)
   \end{equation}

   In this part, we utilize the statefinders $S_{3}^{(1)}$ and $S_{4}^{(1)}$ to diagnose the $f(R)$ model. Using Eqs.(\ref{28})-(\ref{30}) and Ref. \cite{Omar Enkhili:2024}, we can obtain,
   \begin{equation}
   	\label{40}
   	S_{3}^{(1)}=\dfrac{\ddot{H}}{H^{3}}-3q-2
   \end{equation}
   \begin{equation}
   	\label{41}
   	S_{4}^{(1)}=-(1+z)\frac{dS_{3}^{(1)}}{dz}+S_{3}^{(1)}-3S_{3}^{(1)}\dfrac{\dot{H}}{H^{2}},etc
   \end{equation}
   where $\dddot{H}$ represents the third time derivative of the Hubble rate of the Universe.
   
 \subsubsection{The growth rate of perturbations}
 
  The fractional growth parameter $\epsilon(z)$, serving as a null diagnostic tool, is defined in the form\cite{Zhang:2009qa}:
   \begin{equation}
   	\label{42}
   	\epsilon(z):=\frac{f(z)}{f_{\Lambda CDM}(z)}
   \end{equation}
   while $f(z)=dln\delta/dlna$, which describes the growth rate of the linear density perturbation,
   \begin{equation}
   	\label{43}
   	f(z)\simeq\Omega_{m}(r)^{\gamma},
   \end{equation}
   \begin{equation}
   	\label{44}
   	\gamma(z)=\frac{3}{5-\frac{w}{1-w}}+\frac{3(1-w)(1-\frac{3}{2}w)}{125(1-\frac{6}{5}w)^{3}}(1-\Omega_{m}(z))
   \end{equation}
   where $w$ either is constant, or varies slowly with time. For
   $\Lambda$CDM model,  $\epsilon(z) = 1$ and $\gamma\simeq0.55$. However, for $F(R)$ model, the evolutions of $f(z)$ and $\epsilon(z)$ in terms of $z$ depart from the $\Lambda$CDM in Fig.\ref{fig:fig6}. Therefore, $\epsilon(z)$ can be jointly employed with the statefinder parameters to establish a composite null diagnostic (CND), $i.e.$ $\left\lbrace S_{n}^{(1)}, \epsilon \right\rbrace$.

   To begin with, in Fig.\ref{fig:fig5}, the evolution of $S_{3}^{(1)}$ with  $z$ for our model, the current-day value of  $\left\lbrace S_{3}^{(1)},\epsilon\right\rbrace $ for the models are represented by circular markers, while the $\Lambda$CDM fixed point equals to  $(1, 1)$ is marked with a star symbol for reference. Arrows denote the evolutionary trajectories of each model. It is evident that within the low-redshift regime, the $f(R)$ model is readily distinguishable from $\Lambda$CDM. When dimensionless parameter $\lambda$ takes different values, their evolution curves present different trajectories. It is obvious that $\lambda=1.24\Lambda$ is the closest to the $\Lambda$CDM  among them.
   
    Furthermore, through  $\left\lbrace S_{4}^{(1)},z\right\rbrace $, we   displays the evolutionary trajectories of our model in the $S_{4}^{(1)}-\epsilon$ plane in Fig. \ref{fig:fig7}, including both the  values of $\left\lbrace S_{4}^{(1)},\epsilon\right\rbrace$ in low-region redshift for this model and the round point $(1, 1)$ corresponding to the $\Lambda$CDM scenario. Obviously,  we have the current values are marked by the round dots. Both our model and CND under consideration can be differentiated from $\Lambda$CDM model quite well.
   
    \begin{figure}[ptb]
   	\centering
   	\includegraphics[width=0.34\textwidth]{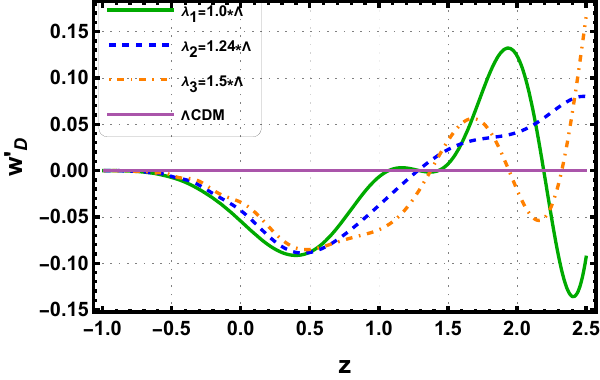}
   	\includegraphics[width=0.3\textwidth]{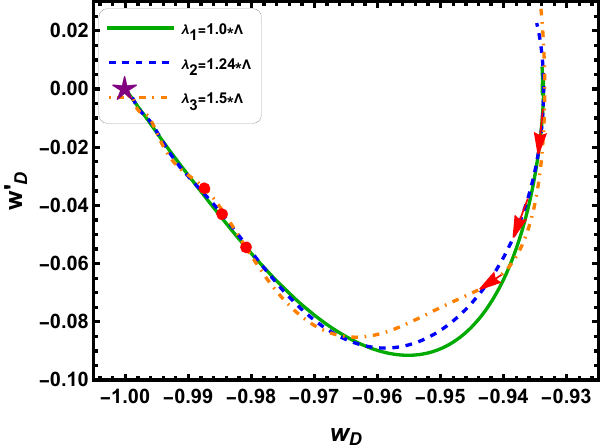}
   	\caption{\small{When $\lambda$ takes different values ($\lambda_{1} =1.0\Lambda$, $\lambda_{2} = 1.24\Lambda$, $\lambda_{3} = 1.5\Lambda$), the  parameter $w'_{D}$ (left plot) versus  $z$ for the polynomial type $f(R)$ model and $w_{D}-w'_{D}$ diagram }}
   	\label{fig:fig8}
   \end{figure}

    \subsection{Analysis by the $w_{D}-w'_{D}$ plane}
    In this part, we introduce another diagnosis $\left\lbrace w_{D},w'_{D}\right\rbrace$ to analyze the cosmological evolution in Ref.\cite{Guo:2006pc}, here prime denotes derivative respect to $x=lna$. Using this relation, we obtain Fig.\ref{fig:fig8}. The  $w_{D}-w'_{D}$ diagnostic is plotted for the $f(R)$ model. The current values are marked by the round dots.  As to the $\Lambda$CDM, $w_{D}=-1$ and $w'_{D}=0$, which is shown as a star for  comparison. The arrows indicate the evolution direction of the model.
    
 \begin{figure}[ptb]
	\centering
	\includegraphics[width=0.32\textwidth]{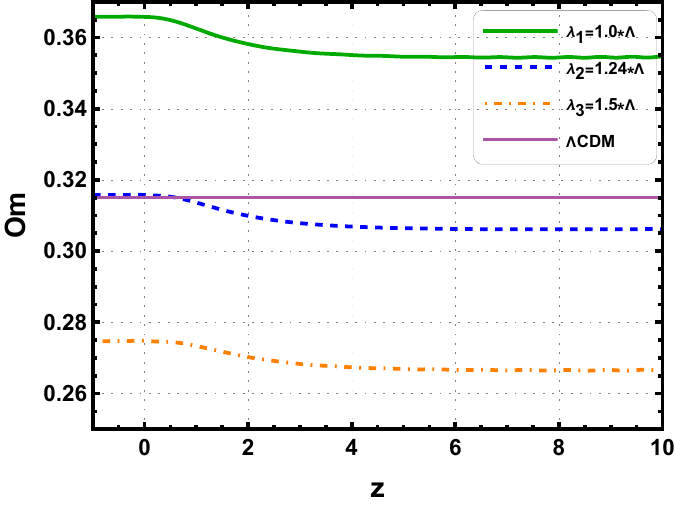}~~~~
	\includegraphics[width=0.32\textwidth]{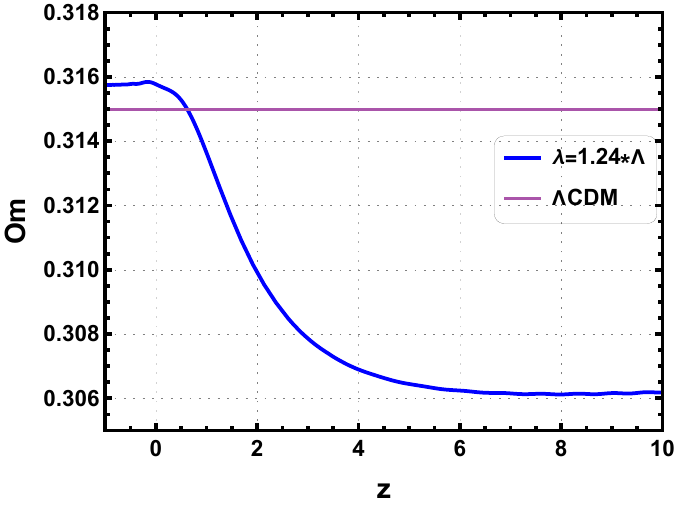}
	\caption{\small{When $\lambda$ takes different values ($\lambda_{1} =1.0\Lambda$, $\lambda_{2} = 1.24\Lambda$, $\lambda_{3} = 1.5\Lambda$), $\eta=0.09$ and $\mu^{2}=1.37\times10^{-67}eV^{2}$, statefinder quantity $Om(z)$ versus  $z$ for the polynomial type  $f(R)$ model. of Eq.(\ref{45}).}}
	\label{fig:fig9}
\end{figure}
   As shown in Fig.\ref{fig:fig8}, the diagnostic images clearly illustrate the dependence on various parameters. Specifically, trajectories associated with  $\lambda=1.24\Lambda$ value  tends to converge more closely to the fixed point of the $\Lambda$CDM, which is (1,0). By examining the evolutionary trajectory of the model, it is evident that our model diverges from the $\Lambda$CDM. Nevertheless, as time progresses, these trajectories are projected to converge back towards the  $\Lambda$CDM .
     
    \subsection{$Om$ diagnostic}
   In this section, we introduce another diagnostic tool, $i.e$  $Om$ diagnostic, which has the advantage of relying on Hubble parameter rather than matter density. The $Om$ diagnostic, first proposed in \cite{V.Sahni:2008,Oliveros:2023ewl}, is defined as an efficient method for
   \begin{equation}
   	\label{45}
   	Om(z)=\frac{h^{2}(x)-1}{x^{3}-1},
   \end{equation}
   where $h(x)=H(x)/H_{0}$, $x=1+z$.
   We know $\Omega_{m}^{(0)}=0.315$ in $\Lambda$CDM model. In Fig.\ref{fig:fig9}, the $Om(z)$ value becomes larger for smaller values of $\lambda$ ,and smaller for larger ones. When $\lambda=1.24\Lambda$, we can clearly extract that range of $Om(z)$ value is $(0.306,0.316)$ from $z=10$ to $z=-1$. Therefore we are able to use this simple method to differentiated our model  from the $\Lambda$CDM paradigm.

     \section{CONCLUSIONS}
    In order to study a new cosmological model, we construct a viable polynomial type  $f(R)$ model in this essay.  With the intention of building on the knowledge from the literature, the background equations were written by using a proper statefinder function, $y_{H}(z)$. By setting initial conditions and combining with the cosmological principle, we derived cosmological parameters such as  $\Omega_{DE}$, $w_{DE}(z)$,  $q(z)$,  $H(z)$ and the statefinder quantities, denoted by $y_{H}(z)$. Evolution images were drawn through a program.  We observe that, at the present time($z = 0$), their values are consistent not only with the observations made by Planck ~2018~ but also with the values that the $\Lambda$CDM  has predicted. (see Table 1.).They explain well the evolution of the Universe. 
    
     Additionally, through SN \uppercase\expandafter{\romannumeral1}a data, we select the theoretical data as against a number of observation date and find there is best fit in between theoretical and observational date.
     Finally, in order to discriminate the models, we introduce various diagnostic tools including statefinder diagnosis pair parameters $\left\lbrace r,s\right\rbrace$ , $\left\lbrace r,q\right\rbrace $ , $w_{D}-w'_{D}$ plane, growth rate analysis, statefinder hierarchy and $Om$ diagnostic. As we expected, the model could be well distinguished from $\Lambda$CDM. As is known to us there is an open problem  which has an ill-oscillatory behaviour in our model. In the future, we need solve this problem and research more viable models.

      \begin{acknowledgments}
      	W. Yang's work is supported by the National Natural Science Foundation of China under Grant Nos. 12547110, 12175096.
      	
      \end{acknowledgments}

\end{document}